\documentclass[prd,aps,floats,tabularx,preprintnumbers,twocolumn]{revtex4}
\usepackage{graphicx, epsfig}

\begin{document}

\baselineskip 12 pt

\title{Eternal Inflation is ``Expensive''}

\author{L. Mersini-Houghton$^\ast$ and L. Parker$^{+}$}

\affiliation
{$^\ast$ Department of Physics and Astronomy, UNC-Chapel Hill, CB{\#}3255, Phillips Hall,Chapel Hill, NC 27599, USA\\
$^{+}$ Department of Physics, University of Wisconsin-Milwaukee, Milwaukee, WI, USA}

\begin{abstract}
{\small}
The discovery of the string theory landscape has recently brought attention to the eternal nature of inflation. In contrast to the common belief that eternal inflation may be a generic feature of most inflationary models, in this note we argue that the suppressed amplitude of perturbations due to adiabatic regularization, together with a fine-tuning constraint on the equation of state of the rare inflating pockets with large fluctuations, render eternal inflation expensive in energy and may make it unlikely to occur.
The energy scales of the eternally inflating pockets have to be very close to the transplanckian regime in order to compensate for the suppression of regularized perturbations.

\end{abstract}

\maketitle

\subsection{\bf Criterion for Eternal Inflation }
The eternal nature of inflation relies on the existence of quantum fluctuations $\delta\phi_q$ that are large enough in some region of space to drive the field $\phi$ up its potential hill. It is believed that inflation is almost always generically eternal\cite{guth,eternal}. In every Hubble time $H^{-1}$ the volume of the initial Hubble sized region grows by $e^3 \simeq 20$. Quantum fluctuations $\delta\phi_q$ around the 
mean field value $\phi_0$ have a Gaussian probability distribution. Conventional
calculations in perturbation theory estimate the average fluctuation at horizon 
exit ($k = a(t) H^{-1}$) to be $\delta\phi_q \simeq H/2\pi$,  where 
$\delta\phi_q^2 \equiv  \Delta_{\phi}^2(k,t)$ is the fluctuation spectrum.
Therefore, the width of their Gaussian distribution is also of order $H/2\pi$. The criterion 
for eternal inflation to switch on is the requirement that in each e-folding (every Hubble 
time) at least one of the $20$ newly created Hubble sized regions has its field 
$\phi = \phi_0 + \delta\phi_q$ fluctuate high up its potential hill, in order to start inflation.

Using conventional estimates this condition translates to \cite{guth} 
\begin{equation}
\delta\phi_q \simeq \frac{H}{2\pi} \ge 0.61 \phi \simeq 0.61 \frac{|\dot\phi|}{H}
\ \mbox{or}\  \frac{H^2}{|\dot\phi|} \ge 3.8
\label{conventfluct}
\end{equation}
An immediate implication of this result is that,when the criterion 
Eq.\ (\ref{conventfluct}) for eternal inflation is met, inflaton density perturbations at 
large scales become very large $\delta\rho / \rho >>1$. Physically, the growth of density
perturbations for eternal inflation is a consequence easily understood by the fact that 
large fluctuations are the mechanism that continually drives eternal inflation and 
density perturbations are proportional to the fluctuations' amplitude squared. 

\subsection{\bf Amplitude of Regularized Perturbations}
It was recently discovered \cite{parker} that the conventional calculation of density 
perturbations differs significantly when adiabatic regularization methods are applied 
to the calculation of fluctuations. The fluctuation spectrum in this case 
becomes \cite{parker}
\begin{eqnarray}
\delta\phi_q^2 &\equiv&  \Delta_{\phi}^2(k,t) \cr &=& \frac{H^2 v^3}{32\pi^2}  
                                    (  4\pi \left | H^{(1)}_{n}(v)  \right |^2 -\cr
            && (m_H^2 + v^2)^{-7/2}(8m_H^6 + 3m_H^4 (3 + 8v^2) + \cr
            && 2m_H^2 v^2 (11+12v^2) +  8(v^4+v^6) ) ).
\label{eq:DeltaSquared}
\end{eqnarray}
where $v=k H^{-1} \exp(-Ht)$,  $n=\sqrt{(9/4) - m_H^2}$,
and $m_H = m/H$. 
Here $H^{1}_{n}(v)$ is a Hankel function of the first kind, and
we are treating $H$ and $m$ as constant to first
approximation during inflation.
Note that $v=2\pi H^{-1}\, / \lambda(k, t)$ is the ratio of the radius of the Hubble 
horizon $H^{-1}$ to the wavelength of the inflaton perturbation
$\lambda(k, t)=2\pi a(t)/k$. For a given value of $k$, as already noted in the
previous section, it is conventional to define the time of ``exit'' of the 
corresponding mode from the Hubble horizon as the time when $v = 1$.  
At that time the wavelength of the inflaton fluctuation has expanded to 
about 6 Hubble horizons. 

When $0<m<0.3 H$ and $v=1$, numerical evaluation of Eq.\ (\ref{eq:DeltaSquared}) shows that:
\begin{equation}
\delta\phi_q \simeq 0.1 m  
\label{eq:smallMassFluctuation}
\end{equation}
This is in sharp contrast to the conventional spectrum quoted in the first paragraph.

\subsection{\bf Can Eternal Inflation Occur?}
In this work we explore the implications of the regularized fluctuations spectrum to the rise of eternal inflation. We can safely approximate Eq.\ (\ref{eq:DeltaSquared}) 
around $v=1$ by 
${\delta\phi}_q\simeq 0.1 m$ where $m^{2} = V''(\phi)$.
Therefore the width of the Gausssian probability distribution is of order $0.1 m$ 
instead of the standard $H/2\pi$. The criterion  $\delta\phi_q \ge 0.61 \phi$, for 
eternal inflation to occur, now becomes

\begin{equation}
0.1 m \ge 0.61 \phi \simeq 0.61 \frac{|\dot\phi|}{H}\ \ 
\mbox{or} \ \
\frac{m^{2} H^2}{|\dot\phi|^2} \ge 37
\label{eternal}
\end{equation}

On the other hand, the flatness condition on the inflaton potential \cite{freese} requires that 
\begin{equation}
\frac{\Delta V}{(\Delta \phi)^4} \ll 10^{-7}
\label{flatness}
\end{equation}
which imposes a fine-tuning of the potential. Here $\Delta V, \Delta \phi$ correspond to the change in the potential and the field from the onset to the end of inflation. 
Eq.\ (\ref{flatness}) constrains the curvature of the potential, $V''\equiv m^{2}$ to be 
about 4-orders of magnitude below the Hubble scale $H$. As can be seen from 
Eq (\ref{eq:DeltaSquared}) the criterion for eternal inflation in Eq.\ (\ref{eternal}), 
which is based on the regularized amplitude of fluctuations, can not be as easily 
satisfied. In contrast to the standard estimate of Eqn.\ref{conventfluct}, now fluctuations 
are suppresed by a factor $m_H^{2}$. This requires the Hubble scale to be
orders of magnitude (given by $(m / H)^{-2}$) larger than previously thought in order to compensate for the suppression factor in the amplitude of perturbations. It turns out that when adiabatic regularization for fluctuations is taken into account, then giving rise to eternal inflation is very costly in energy, since its criterion is met only for 
$V \geq 0.1 M_p^{4}$. This condition on $V$ is 4 orders of magnitude 
larger than the conventional estimate for eternal inflation to occur.

One can argue that although the amplitude of perturbations is suppressed as given by Eq.\ (\ref{eq:DeltaSquared}), there may always be some rare very large fluctuation that takes us to sufficiently high energies $H^2 =(8\pi G/3) V \ge 0.7 M_p^{2}$ which satisfy the condition of Eq.\ (\ref{eternal}), thus giving rise to inflation in that region. 
Even when these rare fluctuations that take the field to 
energy scales in, or very near, the transplanckian regime arise, inflation may not be 
easy to achieve because of constraints on the equation of state of the 
total energy density $w$. In order to get a sense of the possible constraints from 
Eq.\ (\ref{flatness}), let us roughly approximate the change in the potential by
\begin{equation}
\Delta V = V - V_0 \approx (1/2) V" \Delta \phi^2 \approx (1/2) m^{2} \Delta\phi^2
\end{equation}
Then we can write the change in the potential as $\frac{\Delta V}{\Delta\phi^4}\approx \frac{m^{2}}{2\Delta\phi^2} \ll 10^{-7}$, where the change on the inflaton field 
$\Delta\phi$ is typically of order the Planck scale $M_p$. Putting the two conditions 
on $V(\phi)$ together, Eqs.\ (\ref{eternal})-(\ref{flatness}), we obtain
\begin{equation}
37 (3 M_p^{2})(1+w) \leq V" \ll 2\Delta\phi^2 10^{-7}\approx  10^{-7} M_p^{2}
\end{equation}
or
\begin{equation}
(1+w) \ll 10^{-8}
\label{w}
\end{equation}
where $w\equiv p/\rho$ is the equation of state of the total energy density content within one Hubble sized volume. 
In obtaining Eq.\ (\ref{w}) we used the covariant energy conservation equation 
$\frac{-\dot\rho}{3H}=(\rho+p)=\rho(1+w)$ and  $|{\dot\phi}|^2 =(\rho+p)$, $m^{2} = V"$.
Eq.\ (\ref{w}) imposes a severe constraint on the pressure 
and energy density of the inflating pocket, namely, $w$ has to be the equation 
of state of a cosmological constant to within one part in $10^8$. 
With the fluctuations in $\phi$ now being of order $M_p$, it is not clear if the
terms in the inflaton energy density and pressure that involve derivatives of the
inflaton field can plausibly be small enough to satisfy the severe constraint on $w$.

Thus, not only does eternal inflation become very ``expensive'' due to the  suppressed
strength of fluctuations which requires a much higher energy scale $H^2$ to compensate 
for it; but, even if  a very large fluctuation that satisfies Eq.\ (\ref{eternal}) arises, 
it is not clear that it will be able to satisfy the constraint on the equation of state $w$.
This may make eternal inflation less likely to occur generically. 

\subsubsection*{\bf Eternal Inflation with a quartic potential}
As a concrete example, consider the case $V(\phi)= 1/4 \lambda \phi^4$.
The condition for a  pocket to undergo eternal inflation is Eq.\ (\ref{eternal}):
${m^{2} H^2}/{(\dot\phi)^2} \ge 37$, with $m^{2}=3\lambda \phi^2$.
Using the field equation in the slow-roll approximation, 
$\dot\phi \approx {-\lambda \phi^3}/{(3H)}$, 
and Friedman equation, 
$H^2 = {8\pi G \rho}/{3}={2\pi \lambda\phi^4}/{(3M_p^{2})}$, 
one finds that the condition for eternal inflation is
\begin{equation}
12\pi^2 \lambda \phi^4 / M_p^4 \ge 37.
\label{eternal2}
\end{equation}
Then
\begin{equation}
V(\phi)\ge  0.078 M_p^{4}
\label{chaos2}
\end{equation}
and
\begin{equation}
\phi \ge  0.75 M_p (\lambda)^{-1/4}.
\label{chaos1}
\end{equation}
With the constraint that $\lambda< 10^{-12}$, this gives $\phi > 750 M_p$, which is an order of magnitude larger than in the conventional treatment. Similarly, the requirement
on $V$ for eternal inflation is 4 orders-of-magnitude larger than in the
conventional treatment. This example illustrates our main point, that it is 
very ``expensive'' to make inflation eternal and difficult to have it below Planck scales. 
The fact the fluctuations are suppressed makes the width of their Gaussian 
distribution very narrow, thereby bringing the energy scales at which eternal inflation 
can occur very close to transplanckian energies.

\subsection{Conclusions}
We have shown here that the mechanism of using fluctuations to drive the inflaton sufficiently up its potential hill in order for inflation to recur in each efolding, in combination with the fact that the magnitude of fluctuations is very 
suppressed \cite{parker}, makes eternal inflation unlikely for subplanckian energy scales.
We can understand the reason why in the example above $V\ge 0.1 M_p ^{4}$  by the fact that the suppressed fluctuations have a very narrowly peaked Gaussian distribution
with width of  $0.1 m$, so the standard deviation for $\delta\phi$ is very small, making it 
very hard to exceed the value $0.61 \phi$. Compensating for the suppressed density perturbations $\frac{\delta\rho}{\rho}$, requires the potential energy $V$ of the inflating pockets to be nearly transplanckian, making eternal inflation ``expensive.''
We have also argued that the larger density perturbations that are required to
drive eternal inflation may make it difficult to satisfy the constraint of Eq.\ (\ref{w})
on the equation of state $w$ in the pockets undergoing eternal inflation. Our conclusion is that the two constraints discussed in this paper, makes eternal inflation more
expensive in energy and may make it unlikely to occur generically.

\end{document}